\documentclass[aps,prc,amsmath,showpacs,superscriptaddress,nofootinbib,twocolumn]{revtex4}
\usepackage{graphicx,epsfig}
\def\roughly#1{\mathrel{\raise.3ex\hbox{$#1$\kern-.75em%
\lower1ex\hbox{$\sim$}}}}
\def\lsim{\roughly<}
\def\gsim{\roughly>}

\begin{document}

\title{An extended equation of state for core-collapse simulations}

\author{M.~Oertel}
\affiliation{Laboratoire Univers et Th\'eories, CNRS/Observatoire de
  Paris/Universit\'e Paris Diderot, 5 place Jules Janssen, 92195 Meudon, France}

\author{A.~F. Fantina}
\affiliation{Institut d'Astronomie et d'Astrophysique, CP-226, Universit\'e Libre de Bruxelles, 1050 Brussels, Belgium}

\author{J.~Novak}
\affiliation{Laboratoire Univers et Th\'eories, CNRS/Observatoire de
  Paris/Universit\'e Paris Diderot, 5 place Jules Janssen, 92195 Meudon, France}

\date{\today}

\begin{abstract}
  In stellar core-collapse events matter is heated and compressed to densities
  above nuclear matter saturation density. For progenitors stars with masses
  above about 25 solar masses, which eventually form a black hole, the
  temperatures and densities reached during the collapse are so high that a
  traditional description in terms of electrons, nuclei, and nucleons is no
  longer adequate. We present here an improved equation of state which
  contains in addition pions and hyperons. They become abundant in the high
  temperature and density regime. We study the different constraints on such
  an equation of state, coming from both hyperonic data and observations of
  neutron star properties. In order to test the zero-temperature versions, we
  perform numerical simulations of the collapse of a neutron star with such
  additional particles to a black hole. We discuss the influence of the
  additional particles on the thermodynamic properties within the hot versions
  of the equation of state and we show that in regimes relevant to
  core-collapse and black hole formation, the effects of pions and hyperons on
  pressure, internal energy and sound speed are not negligible.
\end{abstract}


\maketitle
\section{Introduction}

Supernovae and hypernovae figure among the most spectacular events observed in
the universe because of the immense amount of energy involved. In general, one
can distinguish between thermonuclear and core-collapse events. Here we shall
be interested in the latter. These occur at the end of the life of massive ($M
\gsim 8 M_{\odot}$) stars: when the iron core exceeds the Chandrasekhar mass a
gravitational collapse is induced. At the center a compact star is formed,
which is a neutron star in the classical gravitational supernova. Depending
among others on the progenitor mass, its metalicity, and rotation, as well a
black hole can be formed. These events are known as hypernovae or
collapsars. For about thirty years, simulations have been performed in order
to explore these events and to answer related questions, for example on the
precise conditions for forming a neutron star or a black hole. The
simulations are extremely complex, since they involve many different
ingredients: multi-dimensional hydrodynamics, neutrino transport, general
relativity and complicated microphysics. Despite all the effort, many unknowns
remain in the simulations, in particular on the engine driving a successful
supernova explosion. Apart from the observations via electromagnetic
radiation, the neutrino and gravitational wave signal could give interesting
information on the models.

The microphysics input for the simulations concerns essentially two domains,
the rates for neutrino-matter interaction and deleptonization, i.e. electron
capture, and the equation of state (EOS). Here we will discuss the latter. It
is not an obvious task to construct an EOS. The main difficulty arises from
the fact that very large ranges of (baryon number) densities ($10^{-10}\
\mathrm{fm}^{-3}\lsim n_B\lsim 1\ \mathrm{fm}^{-3}$), temperatures
($0<T\lsim150$ MeV) and hadronic charge fractions ($0< Y_q = n_q/n_B \lsim\,
0.7$) have to be covered. $n_q$ here denotes the total hadronic charge
density, which in many cases is just given by the proton density. Within this
range the characteristics of nuclear matter change dramatically, from an ideal
gas of different nuclei up to uniform strongly interacting matter, containing
in the simplest case just free nucleons but potentially other components
such as hyperons, nuclear resonances or mesons. Even a transition to
deconfined quark matter cannot be excluded. Although there is a large variety
of EOSs available for cold dense matter relevant for the description of
neutron stars~(see for example~\cite{Lattimer06} and references therein), at
present, only a few hadronic EOSs exist which are commonly used in core
collapse simulations, where temperature effects play a crucial role. There is
the one by Hillebrandt and Wolff \cite{hillebrandt1984}, used by some groups
performing supernova simulations, that by Lattimer and
Swesty~\cite{Lattimer91} as well as that by H. Shen \textit{et
  al.}~\cite{Shen98}. The two latter, publicly available, are most commonly
used in core-collapse simulations. They use different nuclear interactions,
but are based on the same limiting assumptions: they take into account
non-interacting $\alpha$-particles, a single heavy nucleus and free nucleons
in addition to the electron, positron and photon gas.

However, in particular at low densities, i.e. below roughly nuclear matter
saturation density $n_B \lsim n_0 \approx 0.16\ \mathrm{fm}^{-3}$
(corresponding to a mass density of about $\sim 10^{14}\ 
\mathrm{g}/\mathrm{cm}^3$)\footnote{We shall work here exclusively with baryon
  number densities, since the baryon number is a conserved quantity, notably
  throughout a hydrodynamic simulation, contrary to the mass density which is
  not conserved. Many codes, for dimensional reasons, work, however, with a
  mass density. The latter can be obtained easily just by multiplying the
  number density by a constant mass, e.g. the neutron mass $m_n$, see the
  comment on this point in \cite{Hempel11}, too.}  the composition of matter
is much more complicated, with a large number of different nuclei. Although
this should not have a large impact on the purely thermodynamical
properties~\cite{burrows1986}, it is important to correctly describe the
composition of matter in order to determine the electron capture rates and
neutrino interactions. Therefore, in the last years, several groups have
started to build EOSs using mainly statistical approaches to improve the low
density part of the EOS (see
e.g.~\cite{Blinnikov09,Heckel09,Hempel09,Sheng11a,Shen11,
  Sumiyoshi08a,Raduta10}). It has been shown that especially the presence of
additional light nuclei can have an influence on the supernova dynamics and
among others on the neutrino
signal~\cite{O'Connor07,Arcones08,Sumiyoshi08a,Hempel11}. We do not discuss
this point in the present paper since we are mainly interested in a high
density and high temperature extension of the EOS.

Up to know, there are less attempts to improve the high density ($n_B \gsim
n_0$) and high temperature ($T \gsim 20$ MeV) part of the EOS, although there
are many indications that probably the physics of the standard EOS is too poor
in this regime, too. First of all, our knowledge about the quantum
chromodynamics (QCD) phase diagram suggests a transition to the quark-gluon
plasma (QGP) within the range of densities and temperatures reachable in core
collapse events, i.e. within the range of our tables. Of course, there are
lots of uncertainties about this phase transition, so that its occurrence
cannot be affirmed, but the possibility has to be kept in mind when employing
a purely nuclear EOS such as the two EOSs by Lattimer and Swesty~
\cite{Lattimer91} (LS EOS) or by Shen et al.~\cite{Shen98} (Shen EOS) up to
densities well above nuclear matter saturation density and temperatures as
high as several tens of MeV. There is some first work including this phase
transition, see \cite{Sagert08}. Secondly, even without thinking about a QCD
phase transition, other forms of (non-nucleonic) matter should appear at high
densities and temperatures. Already for a long time for cold EOS used for
neutron star models, hyperons, pions and kaons have been considered. At
temperatures above about 20 MeV, this point becomes even more crucial. This
has been confirmed by the first attempts to include hyperons and pions in the
Shen EOS~\cite{Shen98} for simulations, see~\cite{Sumiyoshi08b, Ishizuka08,
  HShen11}. The effect of these high density and high-temperature extensions
of the EOS on the simulations is not negligible, see
e.g.~\cite{Nakazato10b,Nakazato10a,Sagert08,Sumiyoshi08b}. Let us in
particular mention that Sagert \textit{et al.}~\cite{Sagert08} found that the
QCD phase transition could induce a second shock wave which in their
simulations leads to a successful explosion. We shall discuss here the
construction of a new EOS including hyperons and pions based on the Lattimer
and Swesty~\cite{Lattimer91} (LS) EOS and the effects on some thermodynamic
quantities important for the simulations.

The paper is organized as follows. In Section~\ref{sec:LS} we briefly recall
the basics of the LS EOS~\cite{Lattimer91} upon which our model is
based. In Section~\ref{sec:model} we present our extension including hyperons
and pions. In the following Section, Section~\ref{sec:constraints} we discuss
the existing constraints we have on the construction of the extended EOS. In
particular we discuss the compatibility with the recent observation of a
neutron star with almost two solar masses~\cite{Demorest10}, claimed to
exclude the existence of additional particles such as hyperons, mesons or
quarks within cold neutron stars. Section~\ref{sec:collapse} gives an
illustration of the usability of the hyperonic EOSs at zero temperature and
beta equilibrium, Sec.~\ref{sec:resultsT} is devoted to a discussion of the
results at finite temperature and we conclude in Section~\ref{sec:summary}.

\section{The Lattimer and Swesty equation of state}
\label{sec:LS}
Let us start the description of our model for the extended EOS with a
description of the original EOS by~\cite{Lattimer91}. We have chosen this EOS
as basis for our work in order to have an approach for the hadronic
interaction at hand different from the attempts to include hyperons in the
Shen EOS, employing a relativistic mean field model~\cite{Sumiyoshi08b,
  Ishizuka08, HShen11}. The motivation is of course that there are large
uncertainties on the hadronic interaction, so that it is interesting to
compare two different types of models. In addition, the LS EOS is one of the
two commonly used EOSs in computational astrophysics, so that a comparison
of existing results in the literature with results from our extended model
should be simplified.

As mentioned above, the LS EOS\cite{Lattimer91} models the matter as a mixture
of one (average) heavy nucleus, $\alpha$ particles, free nucleons, electrons,
positrons and photons. Electrons and positrons are treated as non-interacting
relativistic gas in pair equilibrium, neglecting electron-screening effects;
photons are treated as an ideal ultra-relativistic gas. Equilibrium with
respect to strong and electromagnetic interactions is supposed, while no
$\beta$ equilibrium is assumed, as expected during core-collapse supernova.

Concerning the nuclear part, the LS EOS follows the works by Lattimer et
al.~\cite{lattimer1978} and by Lattimer and Ravenhall ~\cite{llpr1985}. Some
simplifications have been made with respect to Refs.~\cite{lattimer1978,
  llpr1985}, e.g., the neutron skin is neglected and a simpler
momentum-independent nucleon-nucleon interaction is employed instead of a
standard non-relativistic Skyrme parameterization.  Within the inhomogeneous
phase at low density, nuclei are supposed to arrange themselves in a body
centered cubic lattice which maximizes the separation of ions.  According to
the Wigner-Seitz approximation, each ion is at the center of a neutral-charged
cell, surrounded by a gas of free nucleons, $\alpha$ and
electrons. Interactions between the outside gas and the nuclei are taken into
account through an excluded volume. Nucleons are treated as non-relativistic
particles; $\alpha$-particles as hard spheres of volume $v_\alpha = 24\
\mathrm{fm^3}$ forming an ideal Boltzmann gas. As the density increases,
nuclei undergo geometrical shape deformations, until they dissolve in favor of
homogeneous nuclear matter above approximately saturation density. The
formation of non-spherical nuclei ("\textit{pasta-phase}") is described by
modifying the Coulomb and surface energies of nuclei, as discussed in
Section~2.8 of Ref.~\cite{Lattimer91}. The phase transition to bulk nuclear
matter is treated by a Maxwell construction between the two phases. The
configuration of matter and the balance between the different phases is given
by the most thermodynamically favorable state, i.e. the one which minimizes
the Helmholtz free energy of the system. This procedure, minimizing the free
energy, guarantees that the LS EOS is thermodynamically consistent.

Let us stress, however, one point concerning the description of the transition
between homogeneous and inhomogeneous matter in the LS EOS. As discussed in
Ref.~\cite{Raduta10}, it is not satisfactory, since for all subsaturation
densities matter can be viewed as a mixture of nuclei and free nucleons with
consequences on the thermodynamic properties. In particular, all thermodynamic
quantities are perfectly continuous. As mentioned already in the context of
the distribution of nuclei in the inhomogeneous phase, our main interest here
is a discussion of the high density and high temperature part, taking a
consistent and commonly used EOS for the remaining part.

\subsection{Characteristics of the Lattimer and Swesty equation of state}

The nuclear interaction in the LS EOS contains several parameters, which have
been chosen to reproduce reasonable values for properties of symmetric 
(i.e. equal number of protons and neutrons) bulk
nuclear matter at saturation density, for details see the original
work~\cite{Lattimer91}. These quantities are related to a power-series
expansion of the energy per baryon around saturation density at zero
temperature and for 
symmetric matter:
\begin{eqnarray}
\frac{E}{A} &=& -B + \frac{1}{18}\, K\, x^2 + \frac{1}{162} \,K'\, x^3 + \dots
\nonumber \\
&& + \beta^2 \left( J + \frac{1}{3}\, L\, x + \frac{1}{18}
\,K_{\mathit{sym}}\, x^2 + \dots \right) + \dots~,
\end{eqnarray}
where $x = n_B/n_0 - 1$ is the deviation of the baryon number density
from saturation and $\beta = (n_n - n_p)/n_B = 1- 2 \, Y_p$ describes
the asymmetry. The properties of the EOS are thereby given by the
values of the coefficients, $n_0, B, K, K', J, L$. Of course, this can
only give an indication on the general behavior of the EOS, since they
are defined at saturation density and for symmetric matter, whereas in
the context of neutron stars and core collapse events very asymmetric
matter at very different densities is encountered.
 
Nuclear experiments give constraints on the properties of the
saturation density, $n_0$, the binding energy, $B$, the
incompressibility, $K$ and the symmetry energy at saturation,
$J$. Typical values for $n_0$ lie in the range $0.15\ \mathrm{fm}^{-3}
< n_0 < 0.17\ \mathrm{fm}^{-3}$ and the binding energy is $15.6\ \mathrm{MeV} < B < 16.2\ \mathrm{MeV}$. The value of $K$, roughly
speaking, determines the stiffness of the EOS, the
higher the value of $K$, the stiffer the EOS. But, as mentioned above,
it is determined at saturation density and for symmetric matter, such
that this interpretation has to be regarded with caution. Nuclear
physics experiments on the breathing modes like the isoscalar giant
monopole resonance give a value for $K$ at saturation density of $240
\pm 10$~MeV \cite{piekarewicz2010}. The obvious error is rather
small, the result is, however, not uncontested. In particular, the
extraction of this value from data on isoscalar giant monopole
resonances depends on the density dependence of the nuclear symmetry
energy, a quantity under intensive debate in recent years. We thus
think that a larger range of values has to be considered. The commonly
assumed range for the symmetry energy is $28\ \mathrm{MeV} < J < 34\
\mathrm{MeV}$. For the other parameters, the skewness coefficient $K'$,
the symmetry energy slope coefficient $L$ and the symmetry
incompressibility $K_{\mathit{sym}}$ data are not really constraining the value.

The original LS EOS~\cite{Lattimer91} uses $n_0 = 0.155\ \mathrm{fm}^{-3}, B =
16.0\ \mathrm{MeV}$ and $J = 28.6\ \mathrm{MeV}$, values in reasonable
agreement with the constraints\footnote{The value of $J$ slightly differs from
  the one given in Ref.~\cite{Lattimer91} ($J = 29.3\ \mathrm{MeV}$), see
  Table 2 in Ref.~\cite{Hempel11} too.}.  With the original routines,
see~\cite{ls_web}, three sets of boundary and Maxwell construction tables are
provided corresponding to three different values of the nuclear
incompressibility modulus, $K = 180, 220, 375$ MeV. Following the above
discussion, the two extreme values for $K$ used in the LS EOS are in principle
disfavored and the preferred parameter set for simulations should be that with
$K = 220$ MeV. We shall, however, keep the two other sets for two reasons. The
first one is purely historical: in many simulations the parameter set with $K
= 180$ MeV has been used, so that for comparison with the existing
literature it is interesting to have this value at hand. The second one is, as
discussed above, that the narrow range, $K = 240\pm 10$ MeV is not
uncontested. In that sense, the range of values of the LS EOS represents an
extreme variation of the nuclear parameter sets, i.e. it can give an
indication about the uncertainties in the simulations, coming from the
uncertainties on the nuclear part of the EOS and we shall in principle keep
all three values.

Finally, for simplicity we assume that the nucleon effective mass is equal to
the bare mass: $m^*=m$. In Lattimer \& Swesty~\cite{Lattimer91}, the (density
dependent) effective mass term is kept in the equations, so that this
assumption can be generalized.  Indeed, mean field theories predict an average
effective mass $m^*/m$ around 0.6-0.8 (see e.g. Bender et
al.~\cite{bender2003} for a review, and references therein). It has been shown
that the inclusion of a temperature dependent nucleon effective mass in
nuclei, coming from dynamical correlations beyond mean field, may affect the
core-collapse dynamics \cite{donati1994, fantina2009}. However, as mentioned
earlier, the details of the nuclear part are not the aim of the present work,
and we shall keep the original version of Lattimer and
Swesty~\cite{Lattimer91}.

Note, however, one minor correction with respect to the original code.  It has
been recognized \cite{lattimertalk2006, Buras05} that the original LS EOS
underestimated the fraction of $\alpha$-particles. The reason is that the
$\alpha$-particle binding energy $B_\alpha$ has to be measured with respect to
the neutron mass, as all other energies.

\begin{figure*}
\centering
\includegraphics[width=12.0cm]{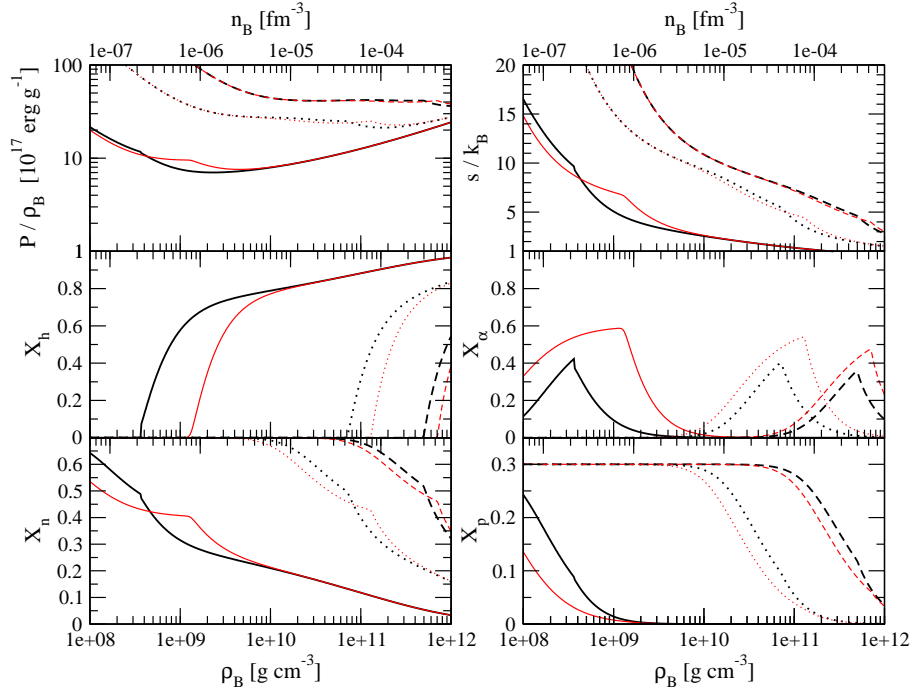}
\caption{(Color online) Pressure, entropy and abundances (nuclei,
  $\alpha$ particles, free neutrons and protons) as a function of
  density, for $K = 180$~MeV, $Y_q = 0.3$ and $T = 1$~MeV (solid
  lines), 2~MeV (dotted lines) and 3~MeV (dashed lines). Thick black
  lines correspond to the results of the original LS routine, while
  thin red lines correspond to the values obtained with the modified
  routine. The differences arise from the correction to the
  binding energy of $\alpha$ particles. We have used here the mass
  density, defined as $\rho_B = m_u n_B$ ($m_u$ being the atomic mass unit), for better comparison with the results of Ref.~\cite{Buras05}.}
\label{fig:ls_balpha}
\end{figure*}

Let us now show the results of this correction to the LS routine. In
Fig.~\ref{fig:ls_balpha} are displayed the abundances, the entropy and
pressure as functions of density, for $K = 180$~MeV, $Y_q =Y_p =Y_e = 0.3$ and
for different temperatures ($T =$ 1, 2, 3 MeV). Note that in this case the
hadronic charge fraction is given by the proton fraction and that it is equal
to the electron fraction, $Y_e = (n_{e^-} - n_{e^+})/n_B$ due to charge
neutrality. These are typical conditions that can be found in a core-collapse
supernova simulation. We observe, as expected, that in the original LS routine
the abundance of $\alpha$ particles is underestimated; as a consequence,
nuclei and free nucleon abundances are higher at a given density. The pressure
of the system (upper-left panel) is not very much affected by the corrections
to the LS routine, since in this density range the contribution from leptons
is the dominant one.  Fig.~\ref{fig:ls_balpha} can be directly compared with
Fig.~4 of Ref.~\cite{Buras05}, where the authors plot the results obtained by
the original LS EOS and by their 4-species (neutrons, protons, $\alpha$ and
$^{54}$Mn as representative heavy nucleus) EOS derived assuming nuclear
statistical equilibrium (NSE). The results obtained by our tables and by the
NSE EOS introduced in Ref.~\cite{Buras05} agree. Nevertheless, differences
have to be noticed, especially for $T = 1$~MeV; this could be explained by the
fact that in their 4-species EOS, the authors assume $^{54}$Mn to be the
representative heavy nucleus, while in LS EOS the mean nucleus varies as a
function of density in order to satisfy the energy minimization
condition. This affects the relative abundances and the macroscopic properties
of the system.

The lowest values of the density in the original routines for the LS EOS is
$n_B = 10^{-6}\ \mathrm{fm}^{-3}$. The physical reason is that, in principle
at low densities and temperatures below roughly 0.5~MeV, an EOS depending only
on temperature, baryon number density and charge fraction to describe matter
in thermodynamic equilibrium is not sufficient and a nuclear reaction network
has to be used. For many purposes, however, a detailed description of matter
in this regime is not necessary, and it is thus interesting to have an EOS at
hand for this regime. Recently, O'Connor and Ott \cite{O'Connor09} have
generated an EOS table in which they employ the LS EOS for densities above the
limiting value of the routines, and, for lower densities, the Timmes
EoS~\cite{timmes1999}, under the assumption that matter is composed of an
ideal gas of electrons, photons, neutrons, protons, $\alpha$-particles and
heavy nuclei with the average $A$ and $Z$ given by the LS EOS at the
transition. We shall follow a slightly different approach, see
Section~\ref{sec:modellow}.
\section{Model for the extended equation of state}
\label{sec:model}
\begin{figure*}
\centering
\includegraphics[width=12.0cm]{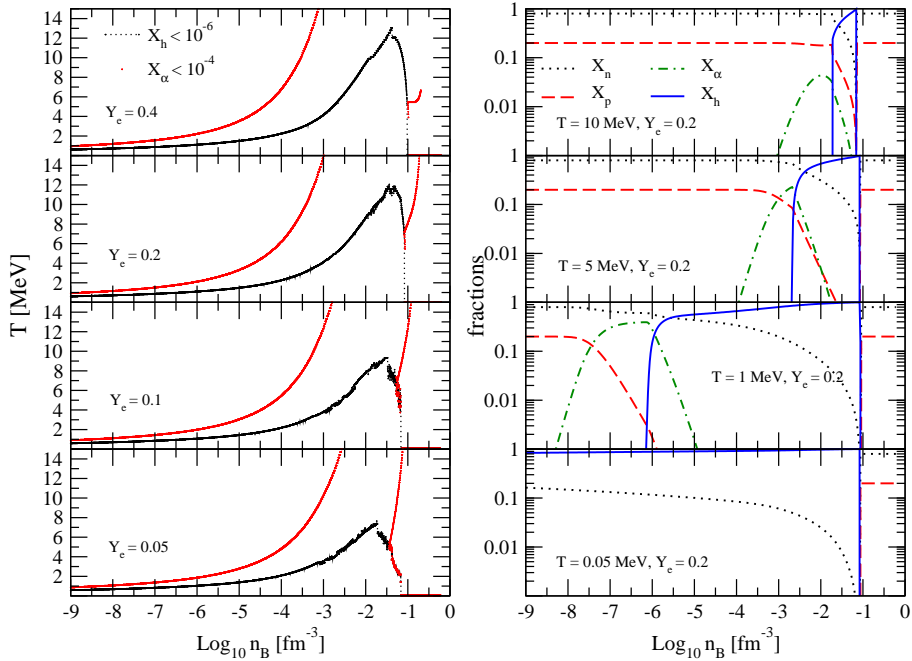}
\caption{(Color online) On the left panels, boundaries between homogeneous and
  inhomogeneous nuclear matter (temperature versus baryon number density) for
  different $Y_q =$ 0.05, 0.1, 0.2 and 0.4 are displayed.  On the right
  panels, mass fractions (free neutrons $X_n$, free protons $X_p$, alpha
  particles $X_\alpha$, and nuclei $X_h$) as a function of baryon number
  density are shown for different temperatures $T = 0.05, 1, 5, 10$~MeV, and
  fixed $Y_p = 0.2$.}
\label{fig:lsshenf1}
\end{figure*}

\subsection{Low density regime}
\label{sec:modellow}
At the densities below the limit of the LS EOS, i.e. $n_B = 10^{-6}\
\mathrm{fm}^{-3}$, at low temperatures matter is composed of a gas of nuclei and
electrons. At temperatures above roughly 1 MeV, nuclei are dissolved
in favor of free nucleons, see Fig.~\ref{fig:lsshenf1}, where the
boundaries between homogeneous and inhomogeneous nuclear matter are
shown. 

In this regime, the densities are so low, that the different particles are
only very weakly interacting and a description in terms of an ideal gas is
completely sufficient. This is the reason why the choice of O'Connor and
Ott~\cite{O'Connor09} to employ the Timmes EOS in this regime (which is
nothing else than an ideal gas of different species) is well justified. We
follow the same idea, the only difference is that we make another choice for
the matter composition, in particular for the nuclei present and for the
matching with the LS EOS at higher densities. O'Connor and Ott take one
average nucleus with $A$ and $Z$ obtained from the average heavy nucleus of
the LS EOS at the transition density. In this way, it is however not possible
to describe a potential variation in $A$ and $Z$ of the mean nucleus
throughout the density range covered by this low density extension of the
EOS. We therefore use a NSE approach, allowing in principle for many different
nuclei to appear. The setup strongly follows the NSE statistical model of
Hempel and Schaffner-Bielich~\cite{Hempel09}, with the only exception that we
do not consider any excluded volume correction because it is not relevant for
such low densities. In particular, we take into account the Coulomb and
temperature corrections to the binding energies of the nuclei in order to
obtain reasonable transition temperatures to homogeneous matter.

Concerning the matching to the LS EOS at higher densities, we have chosen a
matching density of $n_B = 5\times 10^{-8}\ \mathrm{fm}^{-3}$. Although the LS
EOS reproduces well the character of matter at this density, i.e. an almost
ideal gas of nuclei, photons and electrons, the change of the EOS induces
small discontinuities in the thermodynamic quantities, due to the different
treatment of the nuclear part. Since the pressure in this regime is dominated
in any case by the electron pressure, the discontinuity in the pressure is
completely negligible. This is not the case for the energy density and this is
the reason why O'Connor and Ott~\cite{O'Connor09} introduce a constant shift
in the energy density. We judge that the discontinuity is small enough, such
that this shift, problematic in a general relativistic framework, is not
necessary.

\subsection{High density regime}
\label{sec:modelhigh}
We have added to the LS EOS pions, muons and hyperons. For the first two, no
interaction has been assumed and they have just been added as a free gas,
satisfying the overall constraint on charge neutrality. There are many works
considering these additional particles in cold neutron star cores, see
e.g.~\cite{Stone10, Chen11, Schulze11, Burgio11, Weissenborn11b, Bednarek11,
  Massot12}. As already mentioned earlier, there is less work in the context
of hot and dense matter in core collapse events, although the possibility of a
delayed collapse to a black hole induced by a transition to hyperonic matter
has been considered in~\cite{Keil95, Baumgarte96} or pions (and to less extent
kaons, too) have already twenty years ago been considered as possible
candidates for the hot and dense matter in supernova cores, see
e.g.~\cite{Mayle93}. The authors of Ref.~\cite{Mayle93} argue that the
presence of pions could increase the temperature of the supernova core and
increase the number of electron neutrinos and thus lead to a higher neutrino
luminosity in favor of a successful explosion. This idea has, however, not
been further pursued since there are uncertainties about the employed
pion-nucleon interaction. It is now commonly assumed, that there is an
$s$-wave $\pi N$ repulsion, reducing strongly the number of pions eventually
present in supernova cores, thus decreasing the effect described
by~\cite{Mayle93}. This repulsive interaction prevents probably pions from
condensing in cold dense matter as discussed for neutron stars, too. It is,
however, known that the pion gas is one of the main components for matter in
heavy ion collisions. The difference is of course, that for heavy ion
collisions much higher temperatures (of the order of 100 MeV or more) are
reached and the baryon number densities are much lower than in the core of
neutron stars. Core collapse events with massive progenitors are situated
somewhat in between, temperatures can reach the order of 100 MeV, but
densities of several times nuclear matter saturation density are
encountered. Here, we mainly discuss the effect of pions on the EOS at
high temperature where the pion gas should be a reasonable
approximation. Obviously, without interaction, we cannot prevent a $\pi^-$
condensate to form below some critical temperature, depending on the density,
but we consider introducing a realistic pion-nucleon interaction for cold
dense matter to be beyond the purpose of the present paper.

Concerning the muons we have to mention the following point. Since lepton
flavor conversion via neutrino oscillations is most probably negligible for
core collapse during the first few seconds~\cite{Raffelt92,Hannestad99}, muon
lepton number is conserved independently of electron lepton number. Therefore
in principle a muon fraction, $Y_\mu = (n_{\mu^-} - n_{\mu^+})/n_B$ should be
added as variable to the EOS, in addition to the electron fraction $Y_e =
(n_{e^-} - n_{e^+})/n_B$, temperature and baryon number density. This would,
however, mean that the simulation codes should evolve muon number, too. For
the moment this has not been done for several reasons. One is of course the
technical difficulty in adding another evolved quantity together with an
additional dimension for the EOS. The second is that one expects the number of
muons to be much smaller than the number of electrons due to the mass which is
higher by about a factor 200, such that the influence of the muons should not
be very important. With these remarks of caution we show in
Sec.~\ref{sec:resultsT}, under the assumption that muons are in thermal
equilibrium, some examples where the densities and temperatures are high
enough so that the effects of muons are not negligible.

The contribution of pions, muons, electrons and photons to the pressure are
obtained from the expression for an ideal gas,
\begin{equation}
p(\mu,T) = \int \frac{d^3 p}{(2 \pi)^3}
\frac{p}{E} \frac{1}{\exp\left[\beta\, (E-\mu_i)\right] \pm 1}~,
\end{equation}
where $\beta = 1/T$ denotes the inverse temperature, $\mu_i$ is the
chemical potential for particle $i$ 
and $E = \sqrt{m_i^2 + \vec{p}^2}$ is the
single particle energy. The upper sign corresponds to bosons and the
lower to fermions.

Hyperons are added by extending the model by Balberg and Gal~\cite{Balberg97} to
finite temperature. This model is a nonrelativistic potential model
with the contribution of the interaction between particles of type $i$ and $j$
to the energy density given by:
\begin{eqnarray}
\varepsilon_{\mathit{pot}}(n_i,n_j) &=& (1-\frac{\delta_{ij}}{2}) \left(a_{ij}
  n_i n_j + b_{ij} t_i t_j n_i n_j \right.\nonumber \\ && \left.+ c_{ij} \frac{1}{n_i + n_j} (
  n_i^{\gamma_{ij}+1} n_j + n_j^{\gamma_{ij}+1} n_i) \right)~, 
\label{eq:epspot}
\end{eqnarray} 
where $n_i$ denotes the baryon number density of species $i$ and the factor
$1/2$ has been introduced in order to avoid double counting for the
interaction between particles of the same type. $t_i$ represents the third
component of the isospin of the respective particle. $a_{ij}, b_{ij}, c_{ij},$
and $\gamma_{ij}$ are the parameters defining the interaction. 

The total baryonic energy
density is given by the sum of the potential energy, the mass energy,
\begin{equation}
\varepsilon_{\mathit{mass}} = \sum_i n_i m_i~,
\end{equation}
and the kinetic energy,
\begin{equation}
\varepsilon_{\mathit{kin}} = \sum_i \frac{\tau_i}{2 m_i}~.
\end{equation}
Note that, in order to remain consistent with the LS EOS, we do not take an
effective baryon mass into account. 

In order to extent the model to finite temperature we did not change
the structure of the interaction, but we only replaced the
expression for calculating the densities by its finite-temperature version,
\begin{equation}
n_i = \int \frac{d^3p}{(2\pi)^3} \frac{1}{\exp\left[\beta (E_i-\mu_i)\right] +
  1}~.
\end{equation}
$E_i = p^2/(2 m_i) + U_i + m_i$ thereby denotes the single-particle energy for
particle $i$. The kinetic energy densities are written in an analogous way,
\begin{equation}
\tau_i = \int \frac{d^3p}{(2\pi)^3} \frac{p^2}{\exp[\beta (E_i-\mu_i)] +
  1}~,
\end{equation}
see Eq.~(8) in Ref.~\cite{Balberg97}, too. This simple approach of course
neglects possible effects of the temperature on the (phenomenological)
interaction. Investigating these effects is, however, beyond the scope of the
present paper.

The single particle potentials $U_i$ are obtained from the energy
density as $\partial \varepsilon_{\mathit{pot}} /\partial n_i$.
The chemical potentials for the different particles are obtained from
the following relation
\begin{equation}
\mu_i = B_i\mu_B + Q_i\mu_q + L_i^e \mu_{l_e} + L_i^\mu \mu_{l_\mu} + S_i \mu_s~,
\end{equation}
with $B_i, Q_i, L_i^{e/\mu}, S_i$ denoting the baryon number, charge, lepton
number and strangeness of particle $i$. Note that we use the relativistic
definition of the chemical potentials, i.e. the particle rest mass is included
in the chemical potential. This is the reason why we have added the rest mass
to the energy $E_i$, too. We assume that the reaction processes involving
hyperons, e.g. the dominant $\Lambda$ production reaction, $n+n \to n + K +
\Lambda$ are in equilibrium. In addition, we assume equilibrium for
strangeness changing (weak) interactions, such that we can take the
strangeness chemical potential $\mu_s = 0$.
  
We have slightly modified the values of parameters for the
hyperon-nucleon (YN) and hyperon-hyperon (YY) interaction with respect to the
work by Balberg and Gal~\cite{Balberg97} in order to
be compatible with current experimental data on hypernuclei, see
Section~\ref{sec:hyperondata}. This model has the great advantage
that hyperons are added on a nuclear interaction which is exactly the
same as in the original LS EOS by \cite{Lattimer91}, so that an
``artificial phase transition'', induced only by matching one nuclear
model to another and which is thus completely unphysical, is avoided.

\section{Choice of the parameters}
\label{sec:constraints}
This section will be devoted to a discussion of the existing
constraints on the choice of the parameters for the extended EOS. We
start with hyperonic data and then have a look on cold neutron stars. 

\subsection{Hyperonic data}
\label{sec:hyperondata}
In contrast to the nuclear data, hyperonic data are extremely scarce, so that
there are large uncertainties on the hyperonic interactions. Starting with the
description of the fundamental hyperon-nucleon ($YN$) interaction, it is a
long way off having reached the same precision as the nucleon-nucleon ($NN$)
interaction, mainly because of the very limited amount of scattering data. On
the theoretical side, chiral effective field theory calculations have been
performed, too~\cite{Polinder06,Haidenbauer07}, improving on the reliability
of the $YN$ potentials, but still being far from giving conclusive results.
First results from lattice QCD simulations of the $YN$ interaction have become
available~\cite{Beane06}, too.

Concerning the properties of hyperons in dense nuclear matter, there are on
the one hand many-body calculations, starting from the fundamental
interaction. In addition to the traditional $G$-matrix
calculations~\cite{Vidana01,Rijken99,Schulze98}, recently a Hartree-Fock
calculation based on a $V_{\mathit{low} k}$ potential has been presented in
Ref.~\cite{Djapo08}. The results are in reasonable agreement between the
different approaches, but large uncertainties remain due to the not very well
known fundamental interaction, see e.g. the discussion in
Ref.~\cite{Djapo08}. It has been shown, too, that the inclusion of hyperonic
three-body forces ($YNN$) do not strongly change the
results~\cite{Vidana10b}. We should, however, mention that there are very
large uncertainties on the hyperonic three-body force. The authors of
Ref.~\cite{Vidana10b} assume an effective phenomenological form similar to the
interaction in the model by Balberg and Gal~\cite{Balberg97} we are employing
here. They, however, limit the strength of the three-body force arguing that
it should be less important than the nuclear one. Here we follow a slightly
different philosophy: we think that the hyperonic three-body force is not
known well enough to put any constraint on it ad hoc. We shall thus limit it
only by data and by neutron star observations, see next section.

On the other hand, single-particle potentials $V_{YN}$ in symmetric nuclear
matter have been extracted from data on hypernuclei. In that way the empirical
value for the $\Lambda N$-potential at saturation density ($n_B = n_0$),
$V_{\Lambda N} \approx -30$ MeV, has been obtained. This value is in agreement
with an analysis of $(\pi^-, K^+)$ inclusive spectra on different target
nuclei~\cite{Saha04} and reproduced by most many-body calculations, so that
it is commonly accepted. Balberg and Gal~\cite{Balberg97} have adjusted their
parameters to this value, too. For $\Sigma^-$, the situation is somewhat
ambiguous. The observation of a $^4_\Sigma \mathrm{He}$ bound
state~\cite{Nagae98} requires an attractive potential, whereas the analysis of
the $(\pi^-, K^+)$ inclusive spectra~\cite{Saha04,Kohno06} indicate a
repulsive potential, possibly up to a value of $V_{\Sigma N} = 100$ MeV at
saturation density. Theoretical many-body calculations show a large variety of
results, too, see e.g.~\cite{Djapo08}, ranging from slightly less attractive
values as the $\Lambda N$ case to strongly repulsive values of up to several
tens of MeV. Balberg and Gal~\cite{Balberg97} adopt two different versions,
one with an attractive potential of the same form as for the $\Lambda$, and
another local potential form giving rise to a repulsive potential. We here
take the form given in Eq.~(\ref{eq:epspot}), but choosing the parameters in
order to obtain a repulsive single-particle $\Sigma N$ potential. Concerning
the $\Xi N$ single-particle potential, less data are available. Only a few
events of $\Xi$-hypernuclei have been observed, so that it is much more
difficult to reliably fix the well depth of the single-particle
potential. Balberg and Gal~\cite{Balberg97} take a range $V_{\Xi N} =
(-20)$-$(-25)$ MeV, whereas newer data indicate a less attractive potential of
$V_{\Xi N} \approx - 14 $ MeV~\cite{Khaustov99}.

Concerning the hyperon-hyperon $YY$ interaction, the situation is rather
difficult. Early experiments interpreted in terms of production of several
double $\Lambda$-hypernuclei indicate a rather strong attractive potential of
the order of $V_{\Lambda\Lambda} \approx -40 $MeV~\cite{Balberg97}. More recent
measurements~\cite{Nakazawa10} are in favor of much lower values,
$V_{\Lambda\Lambda}\approx -10 $ MeV. For other hyperons, no data are
available. We shall take, in view of the faint knowledge, either no
$YY$-interaction at all or a universal $YY$ interaction with different values
for the well depth. For the isospin dependent terms we follow Balberg and
Gal~\cite{Balberg97}.

\subsection{Neutron stars}
\label{sec:neutronstars}
\begin{table*}
\begin{center}
\begin{tabular}{|l|c|c|c|c|c|c|}
Name & $K$ [MeV]& $M_{\mathit{max}}/M_\odot$ & $V_{\Lambda N}$ [MeV] & $V_{\Sigma N}$ [MeV]& $V_{\Xi
  N}$ [MeV] & $V_{YY}$ [MeV] \\ \hline \hline
180BG & 180 & 1.15 & -28.2 & 16.8 & -24.3 & 0.0 \\ \hline
220BG & 220& 1.70 & -26.6& 28.5 & -22.8 & -38.0 \\ \hline
220g2.8& 220& 1.93 & -29.6& 65.7 & -23.0 & - 55.1 \\ \hline
220g3 & 220& 1.95 & -26.8 & 73.0 & -15.3 & - 10.3 \\ \hline
220pm & 220& 1.94 &  -26.8 & 24.1 & -24.5 & -10.3 \\ \hline

\end{tabular}
\caption{Properties of the different equations of state discussed in the
  text. The cold neutron star maximum mass is given for a non-rotating
  spherical star. }
\label{tab:properties}
\end{center}
\end{table*}
In the center of neutron stars densities of several times nuclear matter
saturation density are reached, so that they present an important test for the
EOS of matter above $n_0$. In contrast to the hot core-collapse environment,
neutron stars older than several minutes can be regarded as cold from the EOS
point of view since the temperature reached is well below 1 MeV. In addition,
$\beta$-equilibrium is achieved and neutrinos can freely leave the system,
so that the EOS in this case is only a function of baryon number density. 

In particular observed masses put constraints on the EOS. There are a number
of precise mass measurements from neutron stars in binary systems, for a
compilation see e.g. Ref.~\cite{Lattimer06}. For a given EOS and a given
central density, the mass and radius of a non-rotating neutron star can be
obtained by solving the equations for hydrostatical equilibrium together with
Einstein's equations. In this case of a spherical star it just gives the
TOV-system. It is known for a long time, that theoretical many-body EOS with
hyperons predict maximum masses of the order $1.4\ M_\odot$ or below,
incompatible with many precisely known neutron star
masses~\cite{Baldo99,Vidana00,Djapo10}. The recent precise measurement of PSR
1614-2230 with a mass of $1.97 \pm0.04\ M_\odot$~\cite{Demorest10} thus
completely excludes those EOSs. The authors of Ref.~\cite{Demorest10} claim
that their data excludes any type of EOS with ``exotic'' contributions,
i.e. other particles than nucleons and electrons. The argument is obvious and
well known: adding new degrees of freedom to the EOS softens it and thus the
maximum mass decreases. This simple argument is, however, only true without
interaction. As has already been shown for EOS with a transition to quark
matter~\cite{Alford07, Weissenborn11}, a repulsive interaction can cure the
problem and allow for neutron star maximum masses of $2\ M_\odot$ or even
above. For hyperons, this seems more difficult, and within the microscopic
approaches, the origin of the necessary repulsion at high densities has not
yet been found. Recently different RMF models have been presented which
successfully reconcile hyperonic matter with PSR 1614-2230,
see~\cite{Hofmann00,Bonanno11,Weissenborn11b, Weissenborn11c, Bednarek11}.
\begin{table*}
\begin{center}
\begin{tabular}{|l|c|c|c|c|c|c|c|c|c|c|c|}
Name & $a_{\Lambda N}$  & $c_{\Lambda N}$ & $a_{\Xi N} $ & $ c_{\Xi N}$ &
$a_{\Sigma N} $  & $ c_{\Sigma N}$ & $ b_{\Sigma N}$& $a_{YY}$ & $ c_{YY}$ &
$b_{\Sigma \Sigma}$ & $\gamma$\\ 
 & MeV fm$^3$& MeV fm$^{3\gamma}$ & MeV fm$^3$& MeV fm$^{3\gamma}$  & MeV
 fm$^3$& MeV fm$^{3\gamma}$ & MeV fm$^3$ & MeV fm$^3$ & MeV fm $^{3 \gamma}$ &
 MeV fm$^3$&
 \\ \hline \hline
180BG &  -505.2 & 605 & -434.4 & 520.1 &10 &175 &214.2 &0 &0 &0 & 4/3  \\ \hline
220BG & -340. & 1087.5& -291.5 & 932.5 & 130.& 300 & 214.2& -486.2& 1553.6&
430 & 2 \\ \hline
220g2.8&-270 &2300 & -170 & 2000 & 500&200 &214.2 &-400 &1500 &430&2.8 \\ \hline
220g3 & -270 & 4000 & -170 & 2900 & 450&250 & 214.2&-90 &1000 &430&3 \\ \hline
220pm & -270 &  4000 & -240 & 3400 &130 &800 &214.2 &-90 &1000 &430 &3 \\ \hline

\end{tabular}
\caption{Parameter values of the different equations of state discussed in the
  text. }
\label{tab:parameters}
\end{center}
\end{table*}

\begin{figure}[h]
\centering
\includegraphics[width = 0.5\textwidth]{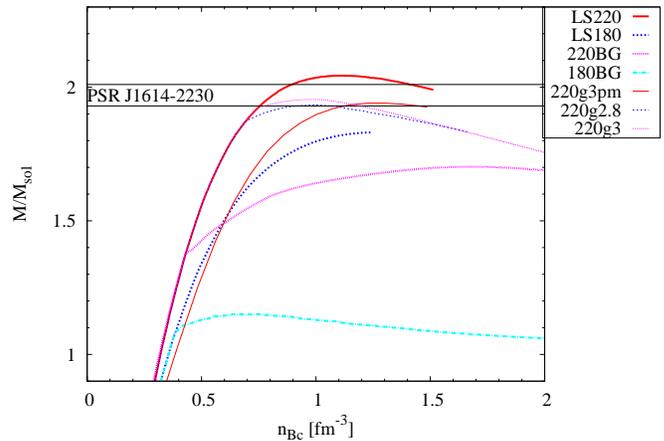}
\caption{(Color online) Gravitational mass of non-rotating spherical neutron
  stars as a function of central baryon number density for the different
  equations of state discussed in the text. }
\label{fig:mnbhyp}
\end{figure}
Here, we take a phenomenological point of view and we choose parameters
for the model by Balberg and Gal~\cite{Balberg97} compatible on the one hand
with hyperonic data and on the other hand with PSR 1614-2230, without looking
for a deeper understanding of the repulsion. Thus, we shall show that it is
possible to reconcile the data, in particular a neutron star mass of $2
M_\odot$ with the existence of hyperons in dense matter, but without answering
the question why theoretical many-body calculations, predicting the
existence of hyperons at densities of about 2-3 $n_0$, cannot reproduce
the $2 M_\odot$ neutron star. The values of the
different single-particle potentials and the maximum mass of a spherical
neutron star for different parameter sets are listed in
Table~\ref{tab:properties}. The first three digits in the name of the
parameter set thereby indicate the value of the incompressibility modulus of
nuclear matter. In Table~\ref{tab:parameters} we list the corresponding
parameters for the hyperonic interaction. The neutron star masses as a function
of central density are displayed in Fig.~\ref{fig:mnbhyp} together with the
constraint from PSR 1614-2230 and the curves obtained with the LS EOS for $K =
180$ and $K = 220$ MeV. Note that pions and muons have only been included for
220pm. 

Parameter set 180BG takes the softest version of the original Balberg and Gal
paper~\cite{Balberg97} with two exceptions: we modified the $\Sigma$
parameters in order to obtain a slightly repulsive $V_{\Sigma N}$ and we
neglected any $YY$ interaction. This parameter set of course gives a far too
low neutron star maximum mass, we include it, however, for comparison. It can
be seen as an extreme case, representing a very (too) soft EOS. Set 220BG
takes the version with the strongest high density repulsion for $YN$ and $YY$
parameters from~\cite{Balberg97}, with, again, one exception: $V_{\Sigma N}$
is chosen repulsive. The other three sets show three examples of parameters
compatible with PSR 1614-2230, still giving reasonable values for the well
depths at saturation density. Let us, however, remark that we did not find any
viable parameter set  with an attractive potential for $V_{\Sigma N}$. For
220g2.8 and 220g3, $V_{\Sigma N}$ is even strongly repulsive, at the limit of
what is compatible with the estimates discussed in
Sec.~\ref{sec:hyperondata}. For 220pm it is weaker, but in that case, no
hyperons are present in cold neutron star matter. The deviation from the LS
EOS case with $K = 220$ MeV, visible in Fig.~\ref{fig:mnbhyp}, thereby arises
only from the presence of pions and muons. 

\section{Collapse of a cold neutron star to a black hole}
\label{sec:collapse}

The goal of this Section is to show the numerical usability of the EOS, by
implementing cold EOSs presented in Sec.~\ref{sec:neutronstars} into a
numerical code and studying the collapse of a cold neutron star to a black
hole. It is not fully relevant to try and measure the effects of the presence
of hyperons or additional particles in this case because it is difficult to
find comparable physical settings: from Fig.~\ref{fig:mnbhyp} it is clear that
the maximal masses with or without hyperons are quite different. Therefore,
initial (unstable) neutron star models used as initial conditions are too
different with different EOSs to provide comparable collapses to a black
hole. The comparison can be performed in the case of a stellar core-collapse,
starting from the same main-sequence massive star initial data and following
then the collapse, bounce, stalling of the shock and collapse of the
proto-neutron star to a black hole, with different EOSs at finite
temperature. As this is beyond the scope of the current paper, it shall be the
subject of a forthcoming study.

\subsection{Transition to a quark phase}\label{sec:quark}
\begin{figure}
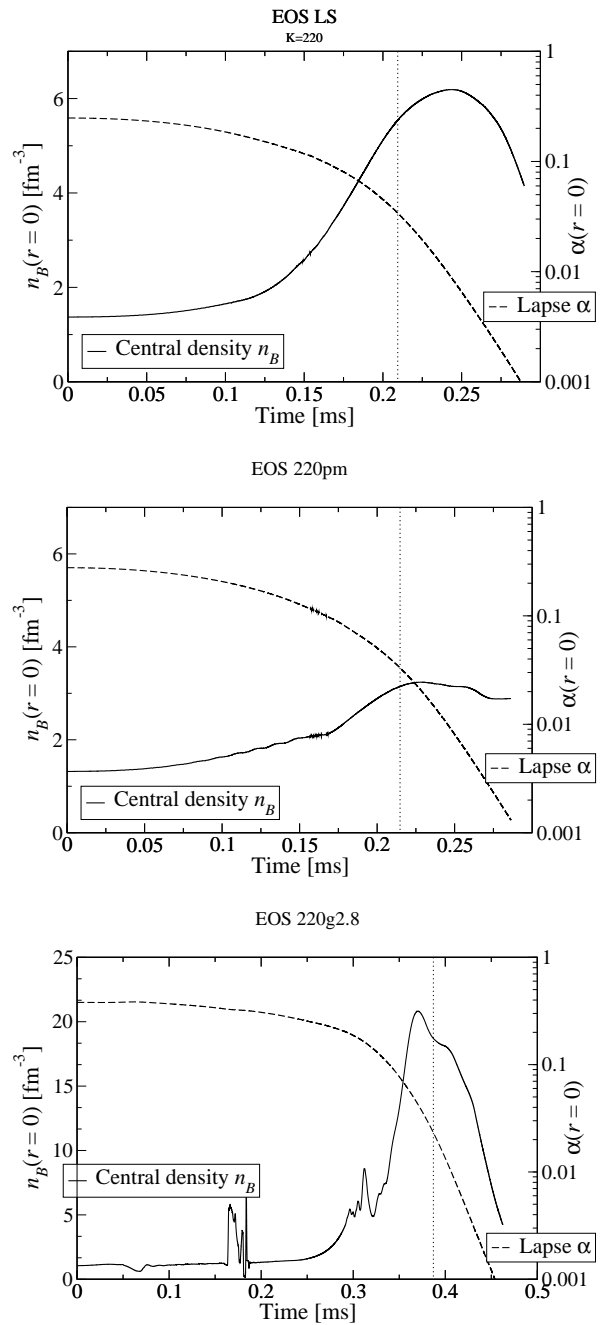

\centering
\includegraphics[width=0.9\linewidth]{collapse_ls220.eps}\\[3ex]
\includegraphics[width=0.9\linewidth]{collapse_220pmq.eps}\\[3ex]
\includegraphics[width=0.9\linewidth]{collapse_220g2.8q.eps}
\caption{Profiles of central density (solid line) and central lapse (dashed
  line) as functions of time during the collapse of a perturbed unstable
  neutron star to a black hole. Top figure was obtained with the LS
  EOS~\cite{Lattimer91} with $K=220$ MeV, the two others by EOSs described in
  Tabs.~\ref{tab:properties} and \ref{tab:parameters}. Vertical dotted lines
  give the time of the formation of the apparent horizon.}
\label{fig:collapses}
\end{figure}

The physical model that we study here is the collapse to a black hole of an
unstable neutron star, i.e. with a central density higher than that
corresponding to the maximal mass. During the collapse, the densities reached
inside the neutron star can be much higher than several times the saturation
density (see Fig.~\ref{fig:collapses} and \cite{gourgoulhon-91}).  As the LS
model for the EOS is non-relativistic, it can in principle allow for a sound
speed greater than the speed of light. With the numerical code we are using
(see Sec.~\ref{sec:numerics} hereafter), such a situation can lead to spurious
oscillations and instabilities destabilizing the whole simulation. This
happens in the LS EOS or in most of the extended EOSs for high densities,
typically beyond 5-10 $\textrm{fm}^{-3}$, that is at more than twenty times
nuclear matter saturation density.  At these high densities, the nature of
matter is anyway far from being well known. It seems rather natural to assume
that there is a transition to a quark matter phase at some density and this is
what we do here. Therefore we have used in our EOS a transition to a simple
model for quark matter, the MIT bag model~\cite{Chodos74} with massless
quarks, implying a supraluminal sound speed. The transition is constructed
using a Maxwell construction. Within this very simple model the density for
the transition can be adjusted by tuning the value of the bag constant. Since
the aim of our paper is not to investigate the transition to quark matter (see
e.g.~\cite{Sagert08} for such a study in the context of core collapse), we try
to push the transition density as far as possible without having superluminal
sound speeds. This means that we choose values of the bag constant much above
commonly used values, leading to appearance of quark matter at a density of
about $n_B = 2\ \mathrm{fm}^{-3}$, the exact density thereby depends on the
specific parametrization of the EOS used.

With the quark phase transition presented here for the
zero-temperature EOS, the sound speed always remains lower than the
speed of light, suppressing all the possible instabilities coming from
superluminal characteristics. In principle this problem could have
been avoided by other prescriptions. For example, a causal EOS, i.e. a
maximally stiff EOS has been used in this context, too. We, however,
think that a transition to quark matter better reflects the physics,
although it should be kept in mind that in this high density region we
do not know much about the real physics and that our quark model
is a very simplistic one.

\subsection{Model setup}\label{sec:numerics}
We use a general-relativistic hydrodynamics code
\textsc{CoCoNuT}~\cite{dimmelmeier-05}, solving the continuity equation and
the relativistic Euler equations for a perfect fluid by means of
high-resolution shock-capturing methods~\cite{font-08}. Einstein equations, in
isotropic gauge and maximal slicing are solved on a different grid with
spectral methods~\cite{grandclement-09}. This code, although able to perform
3-dimensional simulations is run in spherical symmetry since we are
only interested here in illustrating that the different EOSs can be successfully
used in simulations of a collapse to a black hole. The EOS is tabulated,
read from a file and interpolated for every grid point. Static, spherically
symmetric, neutron star initial data are obtained with the same EOS as the one
used for the evolution, and with the same gauge, using the non-rotating
version of the code described in~\cite{lin-06} and the \textsc{lorene}
library~\cite{lorene}. The initial star is chosen to be a spherical
configuration on the unstable branch, with a gravitational mass decreasing if
one increases central density. An unstable neutron star is not bound to 
collapse to a black hole: it can expand in order to ``migrate'' to the stable
branch, \textit{i.e} reach a lower density with the same baryon mass, such
that it is stable with respect to radial oscillations (see
also~\cite{cordero-09}).

The star is then perturbed by amplifying the radial density profile by one
percent. This procedure ensures that the star will collapse to a black hole
and not migrate to the stable branch. The standard picture of such a numerical
model is that the star collapses until general-relativistic effects become
dominant (see e.g.~\cite{gourgoulhon-91} for a complete
description). Among these is the ``frozen-star'' effect which comes from our
choice of time gauge (maximal slicing), avoiding the appearance of a central
singularity. It implies that several quantities do no longer evolve near the
center of the star, whereas some of the metric coefficients develop huge
gradients, limiting the simulation in time. One sign that evolution is frozen
near the center of the star is given by the fact that one metric coefficient,
the lapse $\alpha$, representing the ratio between the physical time measured
by the Eulerian observer and the coordinate time, is tending toward zero. The
second general relativistic effect is the formation of an apparent horizon at
a finite distance, growing further outward until including all the neutron
star matter. This is an evidence of the formation of a black hole. The
2-surface defining the horizon is tracked in our code by an apparent horizon
finder~\cite{lin-07}, which enables us to compute the baryon mass inside the
black hole, too.

\subsection{Results of simulations}\label{sec:collapse0}
\begin{table}
\begin{center}
\begin{tabular}{|r|c|c|c|c|c|}
EOS & $t_{\rm AH} [ms]$  & $t_{99} [ms] $ & $n_B^{\rm max}$ [fm$^{-3}$]& $\Delta M_B$ &
$\Delta M_g$\\
 \hline \hline
LS180 & 0.223 & 0.23 & 18.6 & $2\times 10^{-6}$ & $4\times 10^{-3}$ \\ \hline 
180BG & 0.188 & 0.216 & 114 & $7\times 10^{-6}$ & $2\times 10^{-2}$ \\ \hline
LS220 & 0.209 & 0.216 & 6.19 & $2\times 10^{-6}$ & $4\times 10 ^{-3}$ \\ \hline
220g2.8& 0.387 & 0.396 & 20.8 & $5\times 10^{-6}$ & $6\times 10^{-3}$ \\ \hline
220g3 & 0.327 & 0.334 & 9.51 & $6\times 10^{-6}$ & $5\times 10^{-3}$ \\ \hline
220pm & 0.215 & 0.22 & 3.24 & $10^{-6}$ & $6\times 10^{-3}$ \\ \hline
\end{tabular}
\caption{Characteristics of the six collapses to a black hole studied
  here. EOS names are detailed in the text, $t_{\rm AH}$ is the apparent
  horizon formation time (since the starting of the collapse), $t_{99}$ the
  time at which $99\%$ of the baryon mass has gone into the black
  hole. $n_B^{\rm max}$ is the maximum central density reached during the
  collapse, $\Delta M_B$ and $\Delta M_g$ are the relative conservations of
  baryon and gravitational masses, respectively.}
\label{tab:collapses}
\end{center}
\end{table}
We have run our code on six different tabulated EOSs displayed in the left
column of Tab.~\ref{tab:collapses}. In this table, LS180 and LS220 stand for
the Lattimer-Swesty EOS with the incompressibility $K=180$ and $K=220$ MeV,
respectively. The properties of the four other EOSs are given in
Tabs.~\ref{tab:properties} and \ref{tab:parameters}. The EOS labeled 220BG in
these tables could not give any reliable result in the simulations, because
too much numerical noise appeared already in the initial data. Neutron star
collapses with any of the six EOSs listed in Tab.~\ref{tab:collapses} would
lead to the formation of a black hole in a time of a few tenth of a
millisecond. Time evolution profiles for the density and the central value of
the lapse $\alpha$ are given in Fig.~\ref{fig:collapses} for EOSs LS220,
220pm and 220g2.8. On each of these plots the central density is increasing,
eventually with some oscillations coming from the focusing of the initial
perturbation, before reaching a maximum value and then decreasing. This decrease is
mostly due to the finite resolution at the star center, as the density should
tend toward a given value (frozen-star picture). This spurious decrease is not
an issue because it appears after (or about the same moment as) the formation
of the apparent horizon and therefore in a region inside the black hole that
cannot influence the matter still falling onto it. This problem can in
principle be cured using the excision technique (removing a neighborhood of
the center, replacing it by boundary conditions~\cite{duez-04}), which we plan
to implement in our code in the near future.

On Fig.~\ref{fig:collapses}, the lapse is decreasing toward zero, as expected,
and the apparent horizon forms sufficiently soon to let most of matter
remaining outside the black hole to enter it. The simulations are ended when
the radial gradients of the gravitational fields become too high to be well
described with spectral methods. As seen from the values of $t_{99}$
Tab.~\ref{tab:collapses}, this happens after all matter has been swallowed by
the apparent horizon and therefore one is left with a static Schwarzschild
black hole, with no evolution outside the horizon. From
Fig.~\ref{fig:collapses} and Tab.~\ref{tab:collapses}, one can notice that the
maximal density reached during the collapse strongly depends on the EOS. In
particular, the EOSs with an incompressibility modulus $K=180$ MeV being
softer, matter is more compressed. The addition of hyperons makes the EOS even
softer and, at the end the collapse with EOS 180BG, it reaches such high densities
that they no longer seem realistic (beyond 100 fm$^{-3}$). Finally,
Tab.~\ref{tab:collapses} also gives some error indicators: $\Delta M_B$ is the
relative variation of the baryon mass (number of particles times their rest
mass), $\Delta M_g$ is the variation of the system's gravitational mass (as
deduced from the asymptotic behavior of the gravitational field). The
conservation of baryon mass is directly imposed by solving the equation for
the conservation of the baryon current, whereas the conservation of
gravitational mass is only an indirect consequence of the solved equations and
of the spherical symmetry (no gravitational waves). Therefore, this last
indicator is a good estimate of the overall accuracy of a run. From all these
results, we can claim that, apart from EOS 220BG and 180BG, the cold EOSs
derived here are suitable for numerical simulations in the demanding model of
the collapse to a black hole.

\section{Results at finite temperature}
\label{sec:resultsT}
Let us now discuss the behavior of the EOS with the different
parametrizations at finite temperature. As mentioned earlier,
simulations of core collapse with massive progenitors show that rather
high temperatures of several tens of MeV and even more at
proto-neutron star densities are reached, see e.g.~\cite{Sumiyoshi08b,
  Perego11}. It is clear that the thermal energy is in favor of the
production of additional particles such as hyperons and mesons or
nuclear resonances. The abundances of hyperons, pions and kaons
measured from heavy ion collisions indeed indicate that they are
produced during the collisions. One should of course insist on the
fact that the conditions in heavy ion collisions are different from
core collapse events: the baryon densities are lower and the
temperatures are probably slightly higher, the timescales are such
that no weak equilibrium for strangeness is achieved and the
difference in neutron and proton densities is much less pronounced
than in core collapse events or in neutron stars. But the results show
clearly that the thermal production of those additional particles is
important.

\begin{figure*}
\centering
\includegraphics[width = 12cm]{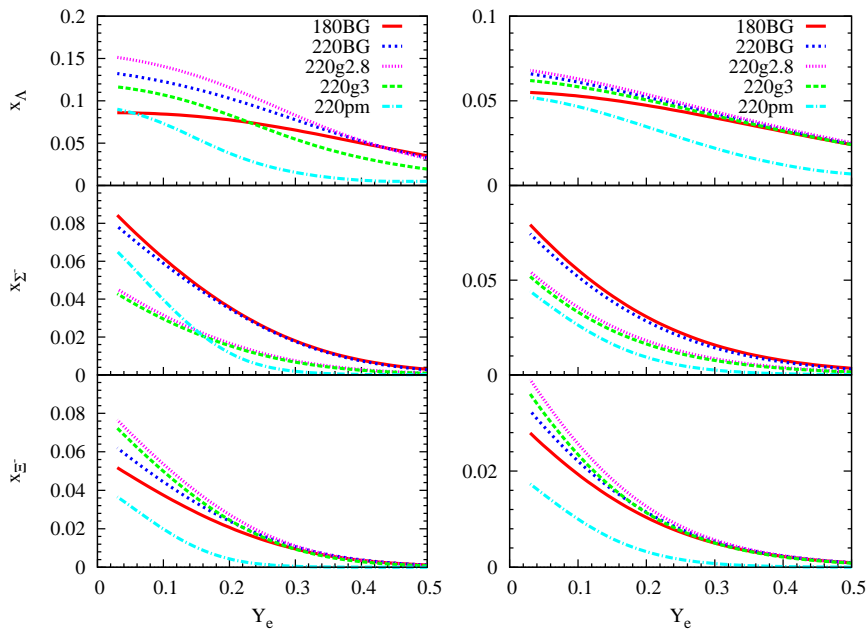}
\caption{(Color online) Fractions of different hyperons as a function of the
  electron fraction at a temperature of 60 MeV and for $n_B = 0.15\
  \mathrm{fm}^{-3}$ (right) and $n_B = 0.3\ \mathrm{fm}^{-3}$ (left)
  corresponding roughly to once and twice nuclear matter saturation
  density. The fractions of $\Sigma^{0,+}, \Xi^0$ are not shown since they are
  below one percent.}
\label{fig:xiye}
\end{figure*}

At finite temperature and without assuming $\beta$-equilibrium, the EOS is a
function of three variables which are generally chosen to be $T, n_B, Y_e$. We
now discuss the properties of the EOS as a function of these variables. Of
course we cannot cover the whole range, but we shall choose some particular
conditions. In Fig.~\ref{fig:xiye} we show the different hyperon fractions as
functions of electron fraction for a temperature of $T = 60$ MeV and two
different densities, $n_B = 0.15\ \mathrm{fm}^{-3}$ and $n_B = 0.3\
\mathrm{fm}^{-3}$. In the simulations of Sumiyoshi et al.~\cite{Sumiyoshi08b}
the first density corresponds to the conditions of a collapse with a $40
M_\odot$ progenitor at bounce at about 10 km from the center. Of course the
exact thermodynamic conditions in a simulation depend on the EOS, so that this
is just to say this is a typical situation within a proto-neutron star after
bounce. High densities are typically reached in the post-bounce phase. We only
show the fractions of $\Lambda, \Sigma^-$ and $\Xi^-$-hyperons since the
fractions of the other hyperons are very small, between one permille and one
percent.

At $n_B = 0.15\ \mathrm{fm}^{-3}$, the $\Lambda$ and $\Xi^-$ fractions show
clearly two groups: the parametrizations without pions and muons and the one
including them. The reason for the different behavior of the EOS if pions and
muons are included is mainly due to the presence of muons. This can be
understood rather easily. Under the present conditions muons are almost
equally abundant as electrons, so that in order to conserve global charge
neutrality, the hadronic charge fraction for a given $Y_e$ is increased by
almost a factor of two with respect to matter without muons. This is clearly
reflected in the curves. Let us stress, however, that the muon fraction is
determined by the degeneracy factor $\mu_\mu/T$ so that it depends on the
ratio of chemical potential and temperature. The curves are calculated
assuming the same chemical potential for electrons and muons. In the case that
neutrinos can freely leave the system, we have $\mu_{\mu^-} = \mu_{e^-} =
\mu_q$, but in the hot proto-neutron star neutrinos are trapped so that there
are nonzero lepton number chemical potentials which are not necessarily the
same for electrons and muons as mentioned already earlier. In general electron
neutrinos are the most abundant ones, so that assuming an equilibrium
Fermi-Dirac distribution the lepton number chemical potential for electrons is
expected to be higher than that of the other leptons. We therefore expect that
in a realistic simulation the chemical potential for muons is lower than that
for electrons reducing the number of muons. It is, however, difficult to
estimate quantitatively the reduction. We can only say that we think that our
results represent an upper limit for the importance of the effects muons can
have on the EOS.

The $\Sigma^-$ fraction shows a stronger dependence on the hyperonic
interaction. Remind that the two parametrizations 220g3 and 220g2.8 contain a
rather strong repulsion for the $\Sigma^-$ single-particle potential, see
Tab.~\ref{tab:properties}, whereas the repulsion is much less strong for the
others. This observation explains the reduction of the $\Sigma^-$-fraction in
these two models with respect to the others. The overall increase in the
$\Sigma^-$-fraction with decreasing $Y_e$ is the well known effect that in
neutron rich nuclear matter negatively charged hyperons are favored. To a
smaller extent the same is observed for the $\Xi^-$, which is, however, less
abundant overall due to its higher mass. Altogether the hyperon fractions
reach maximum values of about 2-8 \%.

\begin{figure*}
\centering
\includegraphics[width = .75\textwidth]{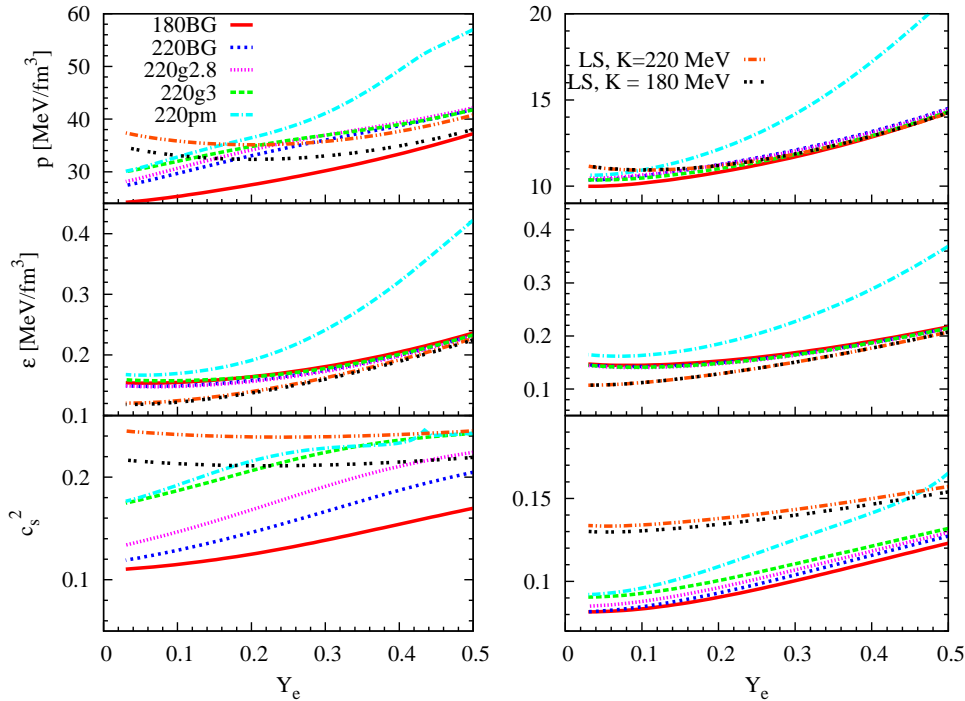}
\caption{(Color online) Thermodynamic quantities as a function of the electron
  fraction at a temperature of 60 MeV and for $n_B = 0.15\ \mathrm{fm}^{-3}$
  (right) and $n_B = 0.3\ \mathrm{fm}^{-3}$ (left) corresponding roughly to
  once and twice nuclear matter saturation density. The upper panels show the
  pressure, the middle ones the internal energy per baryon with respect to the
  proton mass (see text for the definition) and the lower ones the sound speed
  squared.}
\label{fig:thermoye}
\end{figure*}

In Fig.~\ref{fig:thermoye} the pressure, the sound speed and the internal
energy per baryon with respect to the proton mass $m_p$ are displayed for the
same densities and temperature. The latter quantity is defined as
\begin{equation}
\epsilon = \frac{\varepsilon}{n_B m_p} -1~,
\end{equation}
with $\varepsilon$ denoting the total energy density. These three quantities
are key ingredients for the hydrodynamic simulations. We show the results
for the different parametrizations of the hyperonic interactions as well as,
for comparison, the LS EOS with K = 180 MeV and K = 220 MeV, too. 

We again observe that the EOS including muons behaves differently from all the
others. The reason has been explained above. In particular, the usual
softening of the EOS from additional degrees of freedom by adding different
types of particles is not seen. The point is that this softening is
overcompensated by the effect of the increased hadronic charge fraction
induced by the presence of muons, see above. For the other EOS including only
hyperons the softening is indeed seen, as expected it is more pronounced at
higher density (left panels). Compared with the purely nuclear EOS $\epsilon$
is higher including hyperons. This is not very surprising neither since the
hyperons are more massive than nucleons and therefore replacing a nucleonic
state with a hyperonic one in general increases the energy density. The
modifications of pressure and energy density due to the presence of additional
particles is reflected in the sound speeds, too. As already
mentioned above, the hyperon fractions increase with decreasing electron
fraction, so that the effects on the thermodynamic quantities increase with
decreasing $Y_e$, too.

At $n_B = 0.15\ \mathrm{fm}^{-3}$ (right panels) the interaction has only
little effect on the thermodynamic quantities, three groups of EOS can clearly
be distinguished: the two LS ones, those with hyperons and the EOS including
in addition pions and muons. The differences arising from the different
interaction between the two LS ones, and between the four hyperonic ones are
only very small. This can be understood since at high temperature and low
density the kinetic energy should dominate and the interaction terms should in
turn be less important. At higher densities, as can be seen from the figures
at $n_B = 0.3\ \mathrm{fm}^{-3}$ (left panels), indeed the influence of the
interaction is stronger. The difference in pressure between the two LS ones is
of the order of 10\% and between the four hyperonic ones of the order of
15\%. It should be stressed, however, that the softest one, 180BG, gives a
maximum mass for a cold spherical neutron star below $1.2\ M_\odot$ and that it
is thus not very realistic, see
Sec.~\ref{sec:neutronstars}. The difference between the three remaining
hyperonic EOS is much smaller.

\begin{figure*}
\centering
\includegraphics[width = .75\textwidth]{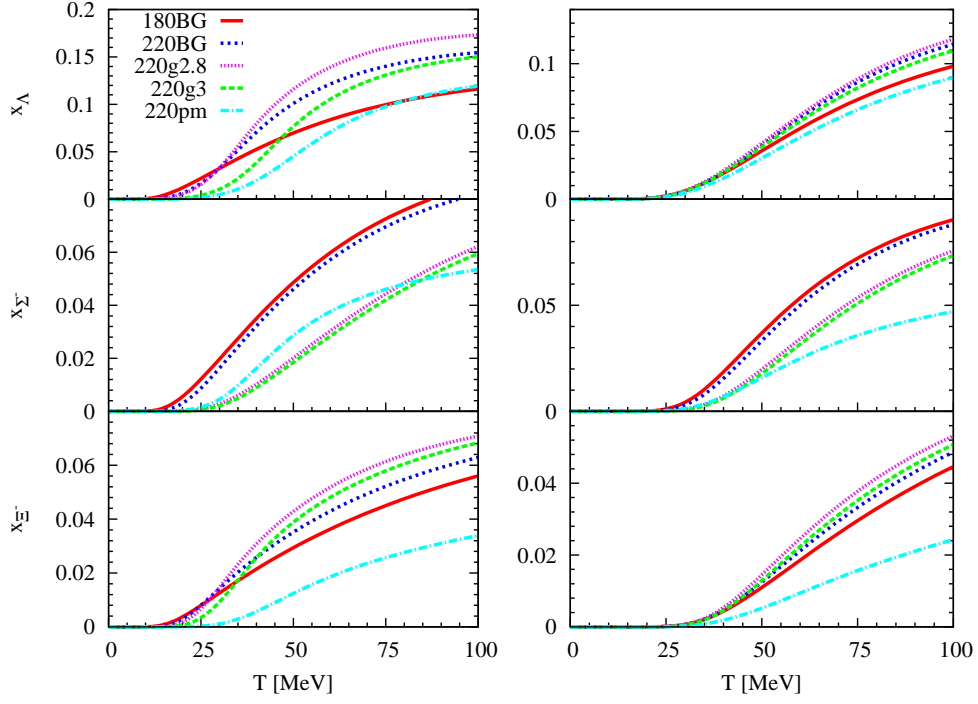}
\caption{(Color online) Same as Fig.~\ref{fig:xiye} but as a function of 
  temperature  for $n_B = 0.15\
  \mathrm{fm}^{-3}$ (right) and $n_B = 0.3\ \mathrm{fm}^{-3}$ (left) and an
  electron fraction of $Y_e = 0.1$. }
\label{fig:xifunct}
\end{figure*}
\begin{figure*}
\centering
\includegraphics[width = .75\textwidth]{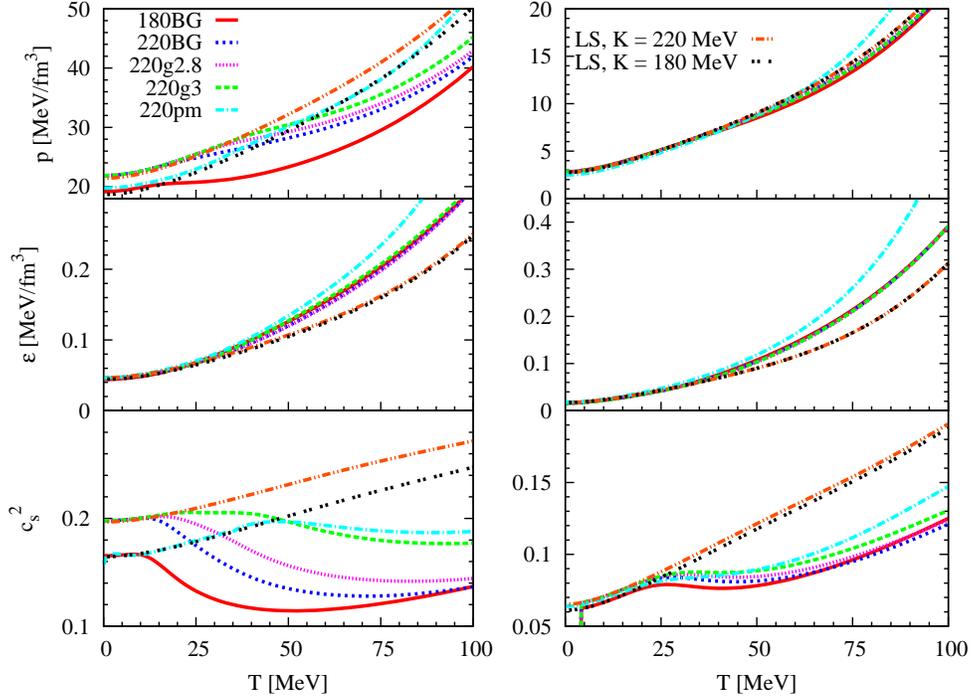}
\caption{(Color online) Same as Fig.~\ref{fig:thermoye}, but as a function of
  temperature for $n_B = 0.15\ \mathrm{fm}^{-3}$ (right) and $n_B = 0.3\
  \mathrm{fm}^{-3}$ (left) and an electron fraction of $Y_e = 0.1$.}
\label{fig:thermofunct}
\end{figure*}

At which temperatures the additional particles in the EOS start to play a
role? In order to answer this question we display in Fig.~\ref{fig:xifunct}
the fractions of $\Lambda, \Sigma^-,$ and $\Xi^-$ and in
Fig.~\ref{fig:thermofunct} the pressure, $\epsilon$ and the sound speed as a
function of temperature. The densities are the same as before, $n_B = 0.15\
\mathrm{fm}^{-3}$ on the right and $n_B = 0.3\ \mathrm{fm}^{-3}$ on the
left. We have chosen a relatively low electron fraction, $Y_e = 0.1$, because
we want to show an upper limit case, i.e. an optimistic estimation of the
effect of the additional particles on the EOS. At the smaller density hyperons
appear at about 25 MeV, independently of the EOS used. The first one to appear
is the $\Lambda$-hyperon. As expected the hyperon fractions rise with
temperature, reaching about 10\% for $\Lambda$ and $\Sigma^-$ and 5\% for
$\Xi^-$ at a temperature of 100 MeV. The differences in the parametrization of
the hyperonic interaction does not induce large differences. Let us make,
however, two remarks. First, the particular features of the EOS with pions and
muons have already been explained and it is thus clear why the hyperon fractions are
systematically lower for this EOS than for the others. Second, for the
$\Sigma^-$-fraction, the two parametrizations by Balberg and Gal, 180BG and
220BG, clearly show a higher $\Sigma^-$-fraction than the other ones. The
reason is again the strong $\Sigma^-$ repulsion in the parametrizations
220g2.8 and 220g3.

At $n_B = 0.3\ \mathrm{fm}^{-3}$, hyperons appear at lower
temperatures, depending on the EOS between roughly 15-25 MeV. The
differences between the EOS are more pronounced than at lower
density. As mentioned earlier, this can be understood from the fact
that, for sufficiently low density, due to thermal effects the kinetic
energy should be dominant rendering the details of the interaction
less important. For this higher density the $\Lambda$-hyperon is
clearly the most abundant one, attaining between 10 and 17 \% at $T =
100$ MeV. The $\Sigma^-$-fraction at this temperature lies between 4
and 8\% and for the $\Xi^-$, slightly less abundant, the fraction reaches
2-6 \%.

The thermodynamic quantities, in particular pressure and sound speed
clearly show the appearance of hyperons, inducing a softening in the
EOS. For the lower density, the modifications in the pressure due to
the presence of hyperons stays, however, relatively small up to $T=
100$ MeV, whereas $\epsilon$ and the sound speed show more important
deviations between the purely nuclear case and the different cases
with additional particles. Again the particle content of the EOS has
more influence on the behavior of the thermodynamic quantities than
the details of the interaction. This is, as already noticed before, not
true at twice this density, where the different parametrizations give different
results for pressure and sound speed. For the energy density the
differences are smaller. 

\begin{figure*}
\centering
\includegraphics[width = .75\textwidth]{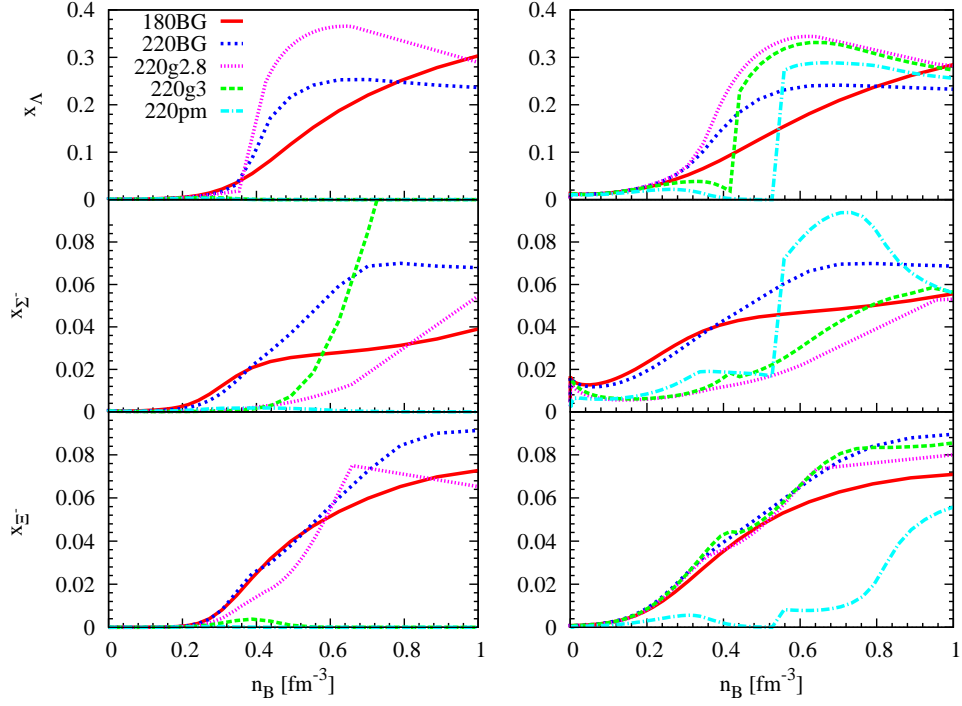}
\caption{(Color online) Same as Fig.~\ref{fig:xiye} but as a function of
  baryon number density for an 
  electron fraction $Y_e = 0.1$ at a temperature of 25 MeV (left) and 40 MeV
  (right).}
\label{fig:xifuncrye10}
\end{figure*}
\begin{figure*}
\centering
\includegraphics[width = .75\textwidth]{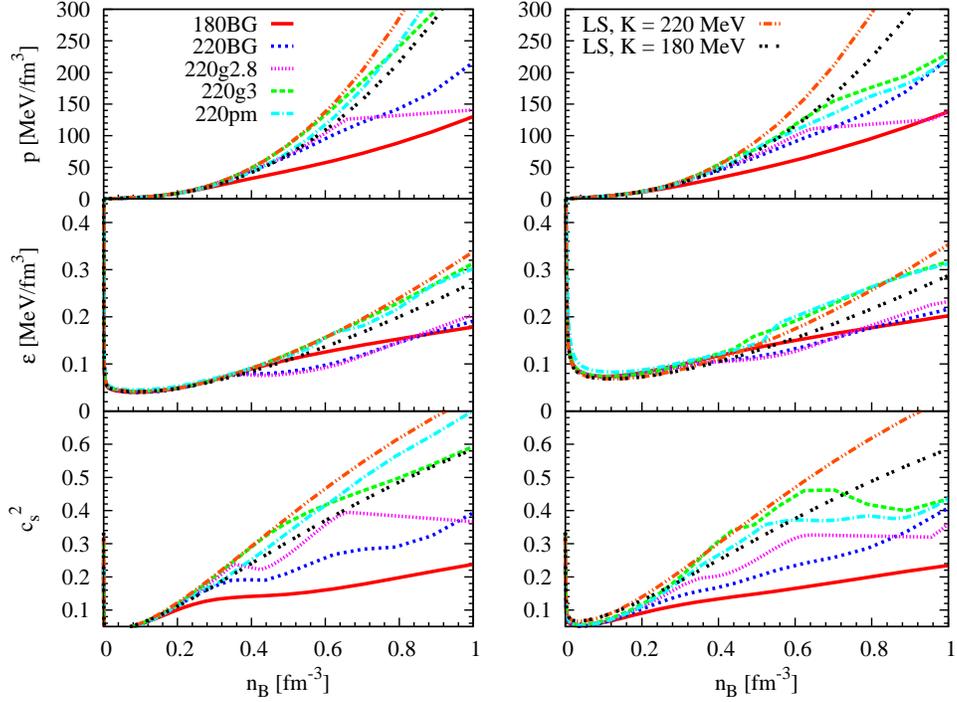}
\caption{(Color online) Same as Fig.~\ref{fig:thermoye} but as a function of
  baryon number density for an 
  electron fraction $Y_e = 0.1$ at a temperature of 25 MeV (left) and 40 MeV
  (right).}
\label{fig:thermofuncrye10}
\end{figure*}

We recover most of the features discussed up to now as a function of density,
too. This can be seen from Fig.~\ref{fig:xifuncrye10}, where the hyperon
fractions are shown and from Fig.~\ref{fig:thermofuncrye10}, where the
thermodynamic quantities are displayed for $Y_e = 0.1$. On the left panels $T
= 25$ MeV, on the right panels $T= 40$ MeV. An interesting point which we have
not seen before because the density has been too low is that at about 2.5$n_0$
a transition takes place, strongly increasing the hyperon fraction and with a
strong effect on the thermodynamics. The $\Lambda$-fraction, for instance, can
in certain models be larger than 30 \%. The thermodynamic quantities, in
particular the pressure, reflect this transition. A thorough discussion of the
thermodynamics of this transition together with a detailed analysis of the
stability is of order but beyond the scope of the present paper.

\begin{figure*}
\centering
\includegraphics[width = .75\textwidth]{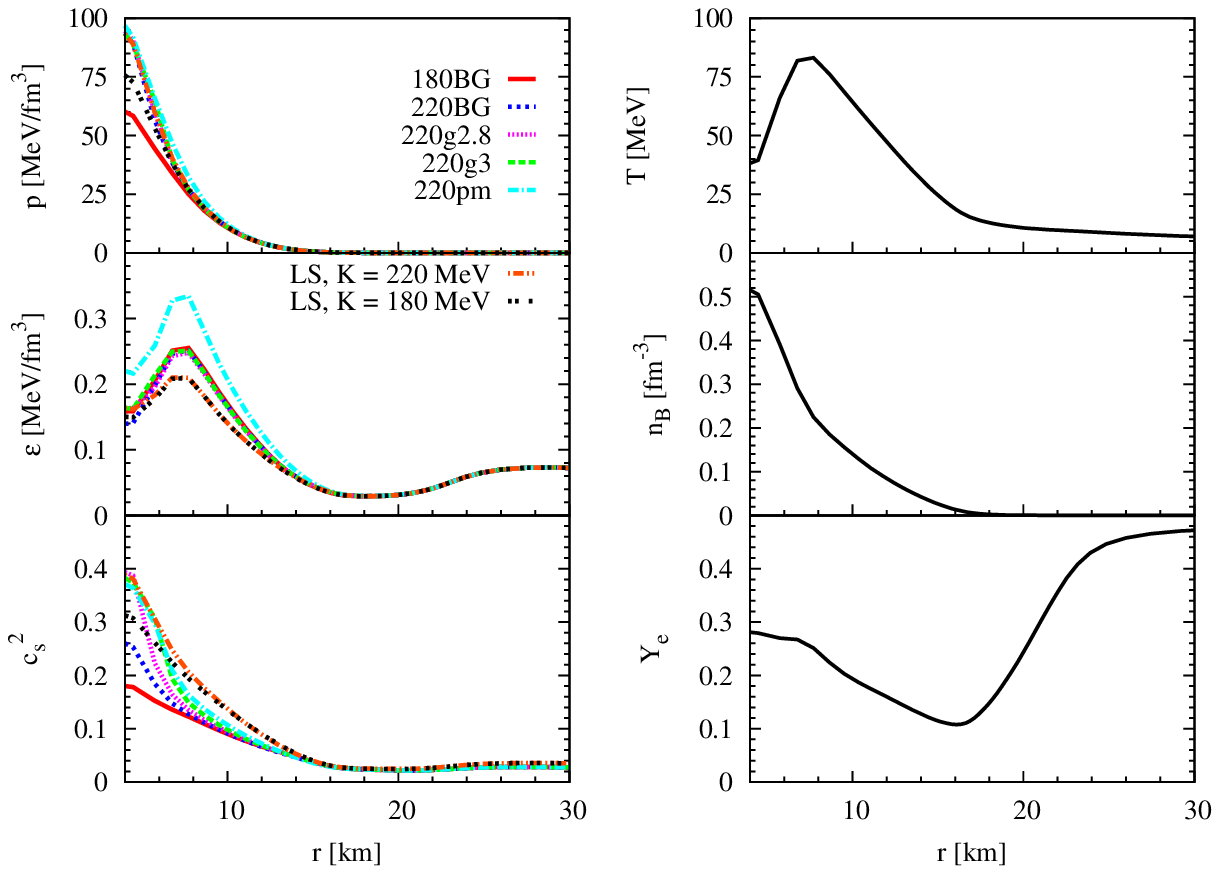}
\caption{(Color online) Pressure, energy density and sound speed (left) for
  different EOS as a function of radius for a proto-neutron star profile
  (right) about
  400 ms after bounce with a 40 $M_\odot$ progenitor. Data (right) are from a
  simulation by M. Liebend\"orfer employing the LS EOS with K = 180 MeV. }
\label{fig:thermoprofile}
\end{figure*}

From the EOS alone we, of course, cannot answer the question whether
the modifications in the thermodynamic quantities due to the presence
of hyperons, pions and muons are relevant for the dynamics of a core
collapse event or a neutron star merger. Here, we do not want to try
to answer this question, but in order to get an idea we compare
the pressure, energy density and sound speed profiles for a hot
proto-neutron star. The data for this profile, thus $T, n_B, Y_e$, are
shown on the right panels of Fig.~\ref{fig:thermoprofile} as a
function of the radius. They are issued by a 1D simulation of the
collapse of a 40 $M_\odot$ progenitor with full Boltzmann neutrino
transport employing the LS EOS with $K = 180$ MeV, see
Ref.~\cite{Liebendoerfer04,Fischer08}, at about 400 ms after
bounce. The left panels show the pressure, $\epsilon$ and the sound
speed as a function of the radius for the different EOS corresponding
to the given values of temperature, baryon density and electron
fraction at this radius. Of course, this procedure does not give
correct proto-neutron star profiles since these depend on the
EOS. However, we think that with this remark of caution in mind, the
comparison of the profiles is interesting and can give hints on the
importance of the modifications in the high density and high
temperature part of the EOS.

Let us first examine the data. The temperature is about 40 MeV at the center,
rising to more than 80 MeV at about 10 km from the center and decreasing then
rapidly to a value between 5-10 MeV. The density is maximal at the center with
a value slightly above $3 n_0$ decreasing to below saturation density at about
10 km from the center. $Y_e$ has a value of about 0.3 at the center. It
decreases until about 15 km, where the value is about 0.1 and rises then,
reaching 0.5 at about 30 km from the center. 

From the behavior of $T, n_B$ and $Y_e$ we would expect sensible
modifications of the thermodynamic quantities only within a radius of
about 15 km from the center of the proto-neutron star. This is indeed
the case as can be seen from the left panels. Let us start with
comparing the two different versions of the LS EOS. In the region very close
to the center, up to roughly 8 km, the pressure and the sound speed
show differences between the two LS EOS. This is understandable since
only in this region the density is high enough to allow for the
differences in the nuclear interaction to play a significant
role. Below saturation density at low temperatures the nuclear EOS is
relatively well constrained so that in this region it would be
surprising to see large differences. For high temperature and low
density the kinetic part becomes dominant, so that no large
differences due to the details of the nuclear interaction are to be
expected. Therefore only the high density region remains where
different nuclear EOSs can show very different behavior. Remember that we
are not interested here in the details of the nuclear composition,
which can have an influence on the thermodynamics and on the
dynamics of a core collapse event too, see ~\cite{Hempel11}. These occur
mainly below saturation density and temperatures below 10 MeV, so
that they cannot be resolved on the scales we are examining here.

Concerning the comparison with the extended EOS, we can see differences up to
a radius of about 15 km. These differences are in general more pronounced than
those between the two version of the LS EOS. The pressure at the center varies
between 60 and 100 MeV$/\mathrm{fm}^3$ from the ``softest'' to the
``stiffest'' EOS. The lowest pressure is obtained for 180BG, the second lowest
for the LS EOS with $K = 180 $ MeV and all the others give values above 90
MeV/fm$^3$. This means that the influence of the additional particles on the
pressure in this proto-neutron star is less important than the value of the
nuclear incompressibility. We of course expect this conclusion to no longer
hold if the central density, only about $3 n_0$ at the stage we are examining
here, increases. And of course, as mentioned before, we have to be careful
since we should recalculate the proto-neutron star profile with having a new
EOS.

For $\epsilon$ we clearly have three groups of EOSs, the LS ones, those with
hyperons and that with hyperons, pions and muons. The differences are most
pronounced at the temperature maximum. This latter point is related to the
fact that the energy density strongly depends on temperature. Moreover, the
profile in $\epsilon$ closely follows the temperature profile. The previous
discussion has already shown why $\epsilon$ has a distinct behavior depending
on the particle content. The sound speed reflects the differences in pressure
and energy density and can vary at the center between roughly $0.45 c$ and $0.6
c$. 

The results for 220pm shown in Fig.~\ref{fig:thermoprofile} have again been
calculated assuming the same chemical potential for electrons and muons. As
discussed above, this represents probably an upper limit for the muonic
effects on the EOS. Here, we have in principle the muon neutrino fraction at
hand. Thus, assuming that the muon neutrinos are in thermal equilibrium,
i.e. that they are described by a Fermi-Dirac distribution, we can determine
the muon lepton chemical potential, and from this the muon chemical
potential. We have computed the profiles with this muon chemical potential
and, as expected, the differences are smaller, but the general trends are the
same and in particular the results for 220pm are still very different from the
other EOSs.

Up to now we have assumed that weak equilibrium with respect to
strangeness is achieved. If, on the contrary, we assume that we have
no weak strangeness changing reactions, which would correspond to
having reaction timescales much longer than the hydrodynamic timescale
of $10^{-6}$~s, strangeness becomes a conserved quantum number. We do
not consider this as a realistic scenario since the timescales
estimated for the relevant processes are of the order of $10^{-6}$ or
below, see e.g.~\cite{Brown92}. We, however, find it instructive to
compare our results with this extreme case. The hyperons have all
negative strangeness, so that populating hyperonic states leads to a
net negative strangeness. Typical production reactions for hyperons
via the strong interaction, for example $n+n \to n + K + \Lambda$, are
strangeness conserving and kaons are produced with positive
strangeness. Thus kaons are the natural candidates for assuring
vanishing net strangeness in thermal equilibrium. However, their mass
of $m_K \approx 500$ MeV is rather high compared with the relevant
chemical potentials and temperatures so that they are not very
abundant~\footnote{Note that we are not discussing here possible
  medium modifications of the kaon properties which could lead to
  higher kaon abundances.} if they are considered
as an ideal gas. This in turn strongly suppresses the hyperon
fractions and therefore the effects on the EOS compared with the more
realistic scenario of strangeness changing weak equilibrium.
\section{Summary} 
\label{sec:summary}
At densities above roughly nuclear matter saturation and temperatures
above several tens of MeV, an equation of state based uniquely on
nucleonic degrees of freedom and electrons is no longer realistic
since many other states will appear.  We have presented here an
extended version of the LS EOS~\cite{Lattimer91} including as
additional particles hyperons, pions and muons intended to improve
on the high density and high temperature part. For zero-temperature
high density matter this question has already been studied for many
years but up to now only very few work exists for finite
temperature. The main problem in this type of exercise is that the
interaction, which is already not well known for nucleons, is even
less known for hyperons. We have adapted here a very simple
phenomenological approach based on the hyperonic model of Balberg and
Gal~\cite{Balberg97}. The parameters of the model have been readjusted
in order to be compatible with available hyperonic data and in
particular the observation of an almost two solar mass neutron star,
PSR 1614-2230~\cite{Demorest10}. Taking these constraints into account
there still remains some freedom, so that we have discussed several
parametrizations of the EOS in order to get an idea of the
uncertainty. The ultimate goal should be, of course, to have a reliable
microscopic approach to hyperonic matter compatible with data, but
awaiting this step, we can, phenomenologically, study the effect of
these additional particles on the thermodynamics of the system. The
results show that key thermodynamic quantities as pressure, energy
density and sound speed are influenced by the additional degrees of
freedom in a non-negligible way. The threshold temperature for the
appearance of hyperons at saturation density lies at about 25 MeV,
depending on the particular model applied. Due to the shift in
the hadronic charge for a given electron fraction, muons seem to
strongly influence the EOS in the regions where they become abundant.  

We have concentrated here mainly on hyperons as additional
particles. Nuclear resonances have not been considered for the
moment. A study of this point is
kept for future work. In another respect our EOS could be improved: we
should not treat pions as a free gas but include interactions. At high
temperatures this should not be very important, but for the low
temperature and high density regime we expect it to have some influence. 

The main application of our EOS should be astrophysical systems, first of
all core collapse events of massive progenitors, collapsing eventually
to a black hole. Neutron star mergers could be another application. We
have demonstrated that our EOS can be successfully used in a numerical
simulation of the collapse of a cold neutron star to a black hole. 
Of course the effect on realistic simulations including finite
temperature has to be tested. 

\acknowledgments 
We would like to thank I. Vida\~na for interesting discussions and
M. Liebend\"{o}rfer for providing us with the proto-neutron star data. This
work has been partially funded by the SN2NS project ANR-10-BLAN-0503 and it
has been supported by
Compstar, a research networking programme of the European Science foundation.
A.F.F. has been also supported by the Communaut\'e fran\c{c}aise de Belgique - Actions de Recherche Concert\'ees, and by the F.R.S.-FNRS (Belgium) via the contract of Charg\'e de Recherches.

\begin{appendix}
\section{Some technical issues on the construction of the EOS tables}
Complete EOS tables, including the extended versions, will be prepared
and made publicly available.  In this Appendix we would like to discuss some
technical issues encountered in the construction of those tables from
the LS original routine. We remind that we have modified the original
routine in two respects: (1) correcting the binding energy of the
$\alpha$-particle (see Fig.~\ref{fig:ls_balpha}); (2) extending the
routine at densities $\lesssim 10^{-6}$~fm$^{-3}$, by extending the
validity of the Maxwell and boundary construction files. This also
permits to verify a correct matching with the low density
EOS. However, some convergence issues at low temperatures and proton
fractions or near critical temperature and density still
remain. Convergence problems manifest in two ways: (1) no solution of
the equilibrium equations, Eqs.~(3.2) in Ref.~\cite{Lattimer91}, is
found (especially at low temperature and electron fraction
\cite{lattimertalk2006}), (2) the solution is discontinuous with
respect to adjacent points in density, temperature, and/or electron
fraction.  In the latter case, the discontinuities appear as: (i) a
rapid changes of the regime (with/without nuclei), which comes from a
cross over the boundaries, (ii) a convergence towards a point far from
the adjacent ones. The first kind of pathology might be due to the
fact that a phase coexistence (and not a phase mixing) is considered
to model the phase transition. The second one might be due to the
sensitivity of the solution with respect to the starting point values
in the minimization routine.

The construction of a table partially overcomes these discontinuities, since,
when looking up the table, an interpolation is done and "critical" points
in-between the grid points are usually avoided. 

Another point to bear in mind when constructing the EOS table is the speed of
sound. This is a crucial quantity in the hydrodynamic simulations, since it
regulates the speed at which the information is propagated and it determines
the time step in finite difference explicit schemes. Throughout the
calculation, due to convergence problems, it might appear that the speed of
sound is either less than zero or superluminal. This is of course non
physical, so for those (rare) points we have recalculated the EOS replacing
the value of the speed of sound with the one obtained in the case of a Fermi
gas (in units of the speed of light) (see e.g. \cite{chabanat1997}):
\begin{equation}
c_s^2 = \frac{(\hbar c)^2}{3(m_e c^2)^2}\ (3\pi n_b Y_e)^{2/3} \ .
\end{equation}
This replacement does of course not concern the region at high
densities where the sound speed becomes superluminal due to the
non-relativistic character of the LS EOS, see Section~\ref{sec:quark}.

We have compared our tables to the O'Connor and Ott ones~\cite{oo_web}.
However, even if there is an agreement in the range where the LS routine is
employed, it is not straightforward to make an exact comparison since the
nuclear parameters that they employ are slightly different from the ones used
in this paper.


\end{appendix}
\bibliography{biblio}

\end{document}